# Exploring hydrogen pipeline costs by considering regional geographical and political characteristics

Bastian Weißenburger [1,2] ✉, Lukas Karkossa [1,3], Annegret Stephan [1] & Russell McKenna [2,4]

Transporting hydrogen using pipelines is becoming increasingly relevant in the energy system, yet current cost estimates typically rely on simplistic approaches that overlook region-specific characteristics, leading to potential miscalculation of costs. Here we examine hydrogen pipeline costs by incorporating regional geographical factors, such as land use, topography and existing infrastructure, together with political-economic factors represented through country-specific weighted average costs of capital. Using a geographic information system-based model, we demonstrate that the regional levelized cost of transportation can vary by up to a factor of three. Comparing our approach with conventional ones based on uniform detour factors in an existing European energy system analysis framework shows substantial deviations in trade flows and highlights the relevance of this work. We provide cost and route data for 4,900 potential global pipeline routes. Our findings yield insights for further research and stakeholders when assessing the economic viability of hydrogen as a competitive energy carrier.

Hydrogen will be a critical energy carrier for hard-to-defossilize sectors in the envisioned net-zero energy system of the future[1–3]. Approximately a quarter of the estimated global hydrogen demand is expected to be traded and transported internationally[4]. The total length of hydrogen pipelines is expected to increase substantially, from around 5,000 km today to 19,000 km by 2030 and 37,000 km by 2035, driven by the rising demand for green hydrogen[1,5]. By 2050, the global network of hydrogen pipelines is projected to reach a total length of 209,000 km (ref. 5). The associated substantial need for infrastructure results in complex challenges in terms of optimal planning. These complex challenges are even underlined by the high uncertainty, given that only 6% of announced projects have reached a final investment decision[6].

Energy system analysis is commonly used for long-term planning of energy systems. Within energy system models[7] such as Enertile[8], PyPSA[9] or TIMES[10], the cost of hydrogen pipeline transport is a crucial parameter. This parameter is often determined heuristically using the pipeline distance and its distance-specific costs.

For new pipelines with unknown routes and distances, current studies tend to estimate the distance involved and then multiply this by a distance-specific pipeline cost[11,12]. Pipeline distance is estimated by combining the Euclidean distance, further referred to as the straight-line distance, with detour factors, which account for any detour the pipeline takes owing to geographical or political constraints[13–16]. Typical values for detour factors on international routes range from 1.20 to 1.40 (refs. 12,15,17,18). To calculate the total investment, this estimated pipeline distance is multiplied by a standard distance-specific pipeline cost—that sometimes distinguishes between offshore and onshore pipelines[11,17].

In practice, the length as well as the specific costs of a pipeline are affected by both geographical and political factors[19–25]. Applying a uniform detour factor overlooks regional characteristics and may bias the results. Figure 1 shows the high variability between actual detour factors on the basis of different sources: the international transmission oil and gas pipelines from the Global Energy Monitor Database (GEM)[26,27] (Fig. 1a) and the proposed European Hydrogen Backbone (EHB)[28] (Fig. 1b). The detour factors are calculated as the ratio of the actual pipeline distance to the straight-line distance, with values higher than 1 indicating a detour compared with the straight-line distance. The interquartile range of the detour factors of these real-world projects ranges from 1.05 to 1.35 for GEM data and from 1.54 to 1.82 for EHB data, spanning from 1.00 to 4.76 in total. While many connections with high values are rather short and therefore less relevant for the following analysis of international transition pipeline routes, the wide range of detour factors highlights the need for individual assessments of each connection and consideration of regional characteristics.

[1]Fraunhofer Institute for Systems and Innovation Research ISI, Karlsruhe, Germany. [2]Chair of Energy Systems Analysis, Institute of Energy and Process Engineering, ETH Zürich, Zurich, Switzerland. [3]Department of Wind and Energy Systems, Technical University of Denmark, Kgs Lyngby, Denmark. [4]Laboratory for Energy Systems Analysis, Paul Scherrer Institute, Villigen, Switzerland. ✉e-mail: bastian.weissenburger@isi.fraunhofer.de

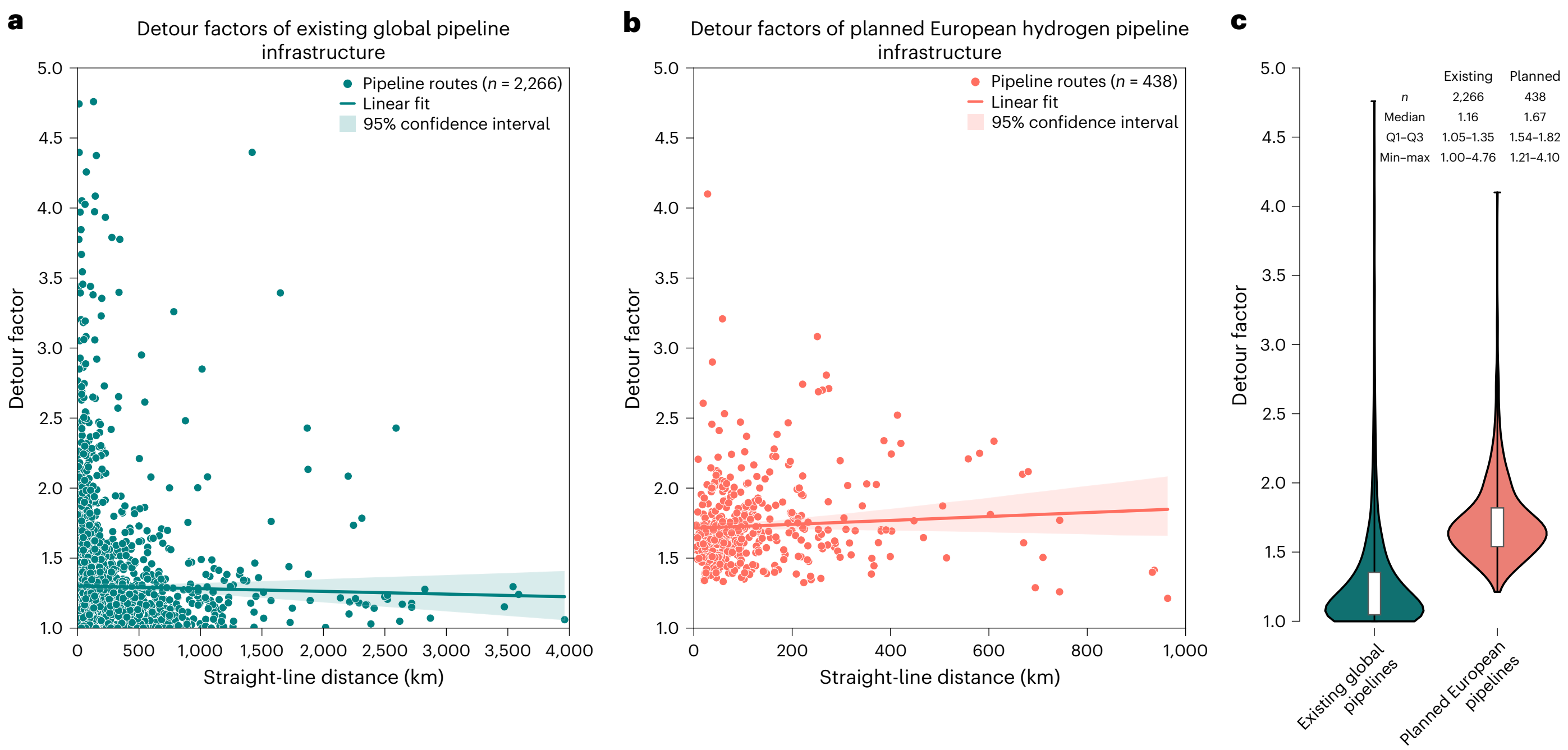

**Fig. 1 | Calculated detour factors for existing and planned pipelines.** **a**–**c**, Comparison of the detour factors for both existing global (**a**) and planned European (**b**) international transmission pipelines from two data sources: Global Energy Monitor Data of existing oil and gas pipelines (GEM)[26,27] and EHB[28] data, respectively. Detour factors, calculated as the ratio of the actual pipeline distance to the straight-line distance, indicate deviations due to geographical or political constraints. The violin plots (**c**) show the distribution of these detour factors, with interquartile ranges between 1.05 and 1.35 for GEM data and between 1.54 and 1.82 for EHB data. This variation underscores the need for individualized assessments of pipeline routes that account for unique regional characteristics.

More specifically, a pipeline's distance-specific cost is influenced by the regional geographical characteristics of its location. Factors such as land use, existing pipeline infrastructure for oil and gas transmission, topography, population density, sea depth, regulatory framework and environmental considerations must all be accounted for[19–23]. While these factors are considered in the detailed planning of pipelines[29,30], they are generally not considered on an international scale, except for sometimes distinguishing between onshore and offshore pipelines[17].

Moreover, extant approaches typically apply a uniform cost of capital[17,31], by using a predefined weighted average cost of capital (WACC), thereby neglecting political characteristics such as individual country risks[31]. However, aspects such as political stability, technical capabilities and the availability of trained workers, permitting procedures and industry structure differ substantially between countries and can affect the cost of (infrastructure) projects substantially[32], indicating the importance of considering different WACCs in energy system models covering different countries[31]. To the best of our knowledge, there is no analysis that considers regional variations on a global and comparative scale.

This Article addresses the identified gap and explores how considering region-specific geographical and political characteristics influences the estimation of hydrogen pipeline transport costs incorporating existing hydrogen pipeline cost analyses in literature[16,28,33–38]. We capture the political dimension by using country-specific WACCs, whose country risk premia reflect economic, political and legal risks. We refrain from modelling hydrogen-specific policy measures to maintain globally comparable, 'level-playing-field' cost estimates. Given the uncertainty regarding future hydrogen infrastructure needs, we analyse potential international pipeline connections worldwide. To do so, we develop a GIS-based model, which determines the optimal pipeline routes between any two points on a global grid with a resolution of approximately 6.5 km × 6.5 km (from ref. 8) and use this to calculate pipeline connections between 264 regions (on the basis of existing work[39]) worldwide.

We find that regional characteristics have a substantial impact on estimated hydrogen pipeline costs, with the levelized cost of transportation (LCOT) differing by up to a factor of three. This difference is also reflected by the calculated route-specific total cost factors, indicating that there is no one-size-fits-all solution, either continentally or globally. While the cost for pipeline transport represents only a proportion of the total supply cost of hydrogen[11,18,40], we show that these can substantially change the results of energy system models. Our dataset consists of about 4,900 global pipeline connections, which we make publicly available. On the basis of our findings, we are confident that considering the geographical and political characteristics for each pipeline produces valuable insights for policymakers and stakeholders. Limitations related to the resolution of our model, the uncertainty of hydrogen pipeline construction costs and the exclusion of existing pipeline repurposing are discussed in Supplementary Note 1.

## Regional variation in hydrogen pipeline costs

Figure 2 shows the routes and the respective LCOT for hydrogen pipeline routes at global scale (Fig. 2a) and their variations globally (Fig. 2b) and for each continent (Fig. 2c–h).

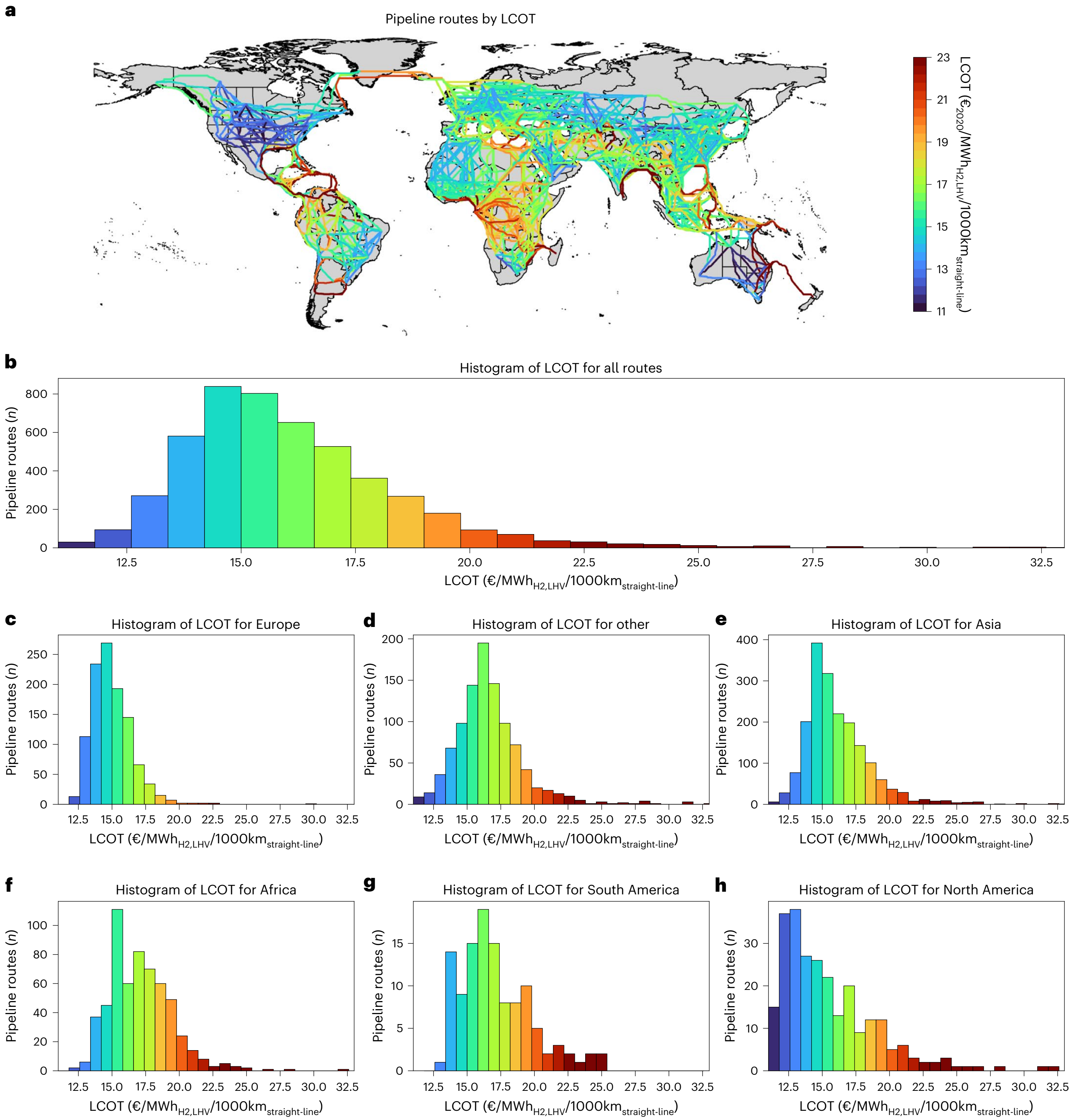

**Fig. 2 | LCOT for global hydrogen pipeline routes. a**, The calculated hydrogen pipeline connections worldwide and the associated LCOT, assuming a uniform diameter for comparability, normalized to the route's straight-line distance. The LCOT varies globally, ranging from 10.9 to 34.2 $€/MWh_{H2,LHV}/1000km_{straight\text{-}line}$, with a threefold cost disparity due to regional differences in geographical and political conditions. **b**, The histogram depicts the global distribution of LCOTs, with most routes falling between 12.7 and 18.6 $€/MWh_{H2,LHV}/1000km_{straight\text{-}line}$, and a standard deviation of 2.9. **c**–**h**, The histograms present the LCOT distribution by continent—Europe (**c**), Asia (**e**), Africa (**f**), South America (**g**) and North America (**h**)—where 'other' (**d**) includes intercontinental routes and Oceania owing to the limited number of connections. Europe and North America have lower average LCOTs of 15.0 and 15.6 $€/MWh_{H2,LHV}/1000km_{straight\text{-}line}$, respectively, while Africa and South America show higher averages of 17.2 and 17.1 $€/MWh_{H2,LHV}/1000km_{straight\text{-}line}$, respectively. The variability within regions emphasizes the inadequacy of applying a single detour factor universally, which is further explored in subsequent sections. A calculation using a uniform WACC of 6% for all countries is shown in Supplementary Fig. 1. Basemap from Natural Earth (https://www.naturalearthdata.com).

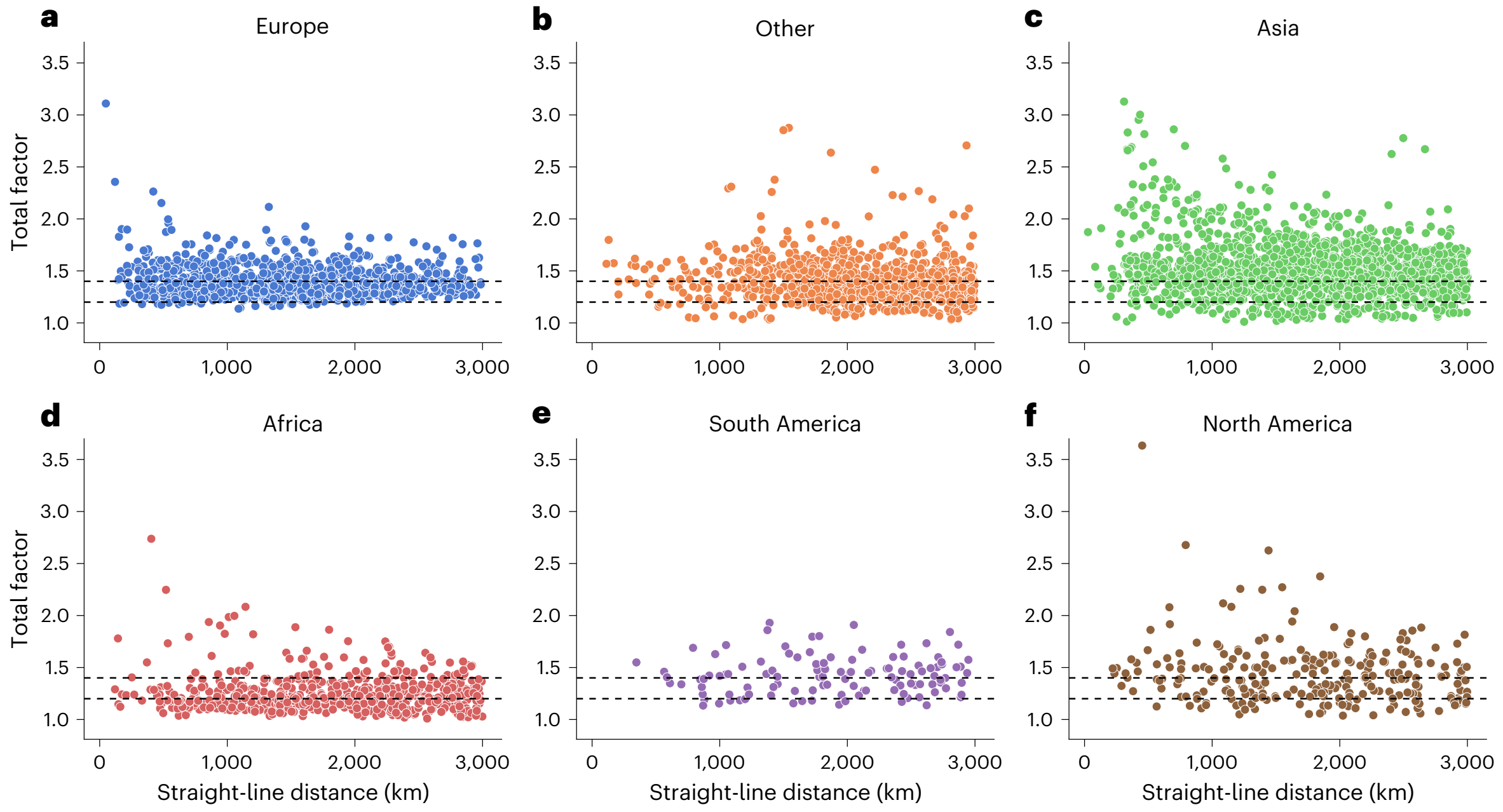

**Fig. 3 | Total factor of each pipeline route by continent. a–f**, The total factor for each pipeline route, categorized by continent. The total factor represents the cumulative increase in pipeline investment due to either extended distances from detours or elevated costs arising from geographic conditions or both. Dashed lines indicate standard literature values, ranging from 1.20 to 1.40. The analysis includes data from Europe (**a**), Asia (**c**), Africa (**d**), South America (**e**) and North America (**f**), with 'other' (**b**) encompassing all intercontinental connections. Oceania is excluded owing to a limited number of routes. Notably, no continent shows a consistent total factor. In Europe and South America, the total factor rarely falls below the lower literature threshold of 1.20, whereas in North America, the factor can reach up to 3.63. A linear descending trend with increasing distances is visible across most continents, except in South America. Five short (<250 $km_{straight\text{-}line}$) routes in China, with total factors of up to 7 as a result of them being forced through densely populated locations, were excluded from the analysis.

Our calculations show that many routes do not follow the straight-line distance. The LCOT ranges from 10.9 to 34.2 euros per megawatt-hour of hydrogen for every 1,000 km of origin-to-destination straight-line distance (€/$MWh_{H2,LHV}$/1000$km_{straight\text{-}line}$, where LHV is lower heating value) globally with a standard deviation of 2.9 €/$MWh_{H2,LHV}$/1000$km_{straight\text{-}line}$, showcasing a nearly threefold increase from the least to the most expensive pipeline. Hence, geography and political factors substantially influence the cost of hydrogen transport in different regions.

Europe, North America and Asia tend to have a lower LCOT, with mean values of 15.0, 15.6 and 16.2 €/$MWh_{H2,LHV}$/1000$km_{straight\text{-}line}$, respectively, whereas Africa's and South America's mean LCOT is higher at 17.2 and 17.1 €/$MWh_{H2,LHV}$/1000$km_{straight\text{-}line}$, respectively. While the distribution of LCOT is relatively narrow in Europe, with a standard deviation of 1.5 €/$MWh_{H2,LHV}$/1000$km_{straight\text{-}line}$, it is wider for the other continents, especially North America, which features some of the routes with the lowest as well as the highest LCOT and a standard deviation of 4.2 €/$MWh_{H2,LHV}$/1000$km_{straight\text{-}line}$. For the remaining regions, the standard deviation is between 2.4 and 3.1 €/$MWh_{H2,LHV}$/1000$km_{straight\text{-}line}$.

This high spread of pipeline costs across the globe and within the individual regions indicates that one factor cannot capture all of the costs well. Therefore, a total factor is defined and investigated in more detail in the next section before being broken down into its two components, detour and cost factor, which represent the route's detour and distance-weighted sum of all geographic cost factors, respectively.

## Influence of detour and geographical characteristics

Figure 3 shows the total factor of different routes for the different continents. The total factor represents the increase in pipeline investments due to detours from the straight-line or due to elevated costs. It is calculated by multiplying each route's detour factor with the cost factor. The cost factor includes the geographical factors influencing construction cost such as topography and land use. Political factors are included in the LCOT through the WACC (see below). Dashed lines indicate the typical range of standard literature values between 1.20 and 1.40.

Most notably, no continent exhibits a consistent total factor. For many routes (48%), even a factor of 1.40 (that is, the higher value used in literature) proves to be an underestimation. In Europe and South America, the total factor rarely (3%) drops below the lower literature threshold of 1.20.

The effects of incorporating geographical and political factors into pipeline route calculations are stronger for some routes than for others. In all continents, factors frequently exceed 1.40, with South America reaching up to 1.93 and North America even up to 3.63. Conversely, factors close to 1.00 are commonly observed in Africa and Asia, less frequently in North America and 'other', but not in Europe and South America, where factors never fall below 1.1.

A descending linear trend with increasing distance is observable across all continents, except South America, possibly owing to fewer shorter routes here compared with other regions.

Figure 4a shows a more detailed analysis of these route differences by further decomposing the total factor into its two components, cost and detour factor, plotting cost factors on the *y* axis and detour factors on the *x* axis. Again, dashed lines indicate the area between 1.20 and 1.40 that is covered by factors from existing literature. Figure 4 also shows the findings from Fig. 3, illustrating that 39% of the routes fall within the established boundaries and 61% lie outside, either above or below these limits. The scatter plot indicates that high total factors are primarily driven by either a high cost factor or a high detour factor but rarely both (dots would have to lie in the upper-right quadrant). Notably, routes with high detour factors often correlate with higher distances compared with those with higher cost factors.

Figure 4b shows four example cases that lie outside the literature range. These comprise a route with a very low total factor (a) and three routes with higher total factors, one resulting from a high cost factor (b) and two from an increased detour factor (c and d). Figure 5a–d shows the exact course of each route in the respective map including the corresponding LCOT per tile in €/$MWh_{H2,LHV}$/$1000km_{straight\text{-}line}$. Route a, connecting Benin to Algeria, has a total factor of 1.02. This route is characterized by both low cost and detour factors. It runs 2,909 km through Western and Northern Africa and only deviates noticeably from the straight line to bypass the W National Park in northern Benin and to align with existing pipeline corridors in Algeria. The cost of this route benefits from favourable geography, primarily barren or savannah regions, which do not incur additional costs, along with low population densities and existing infrastructure.

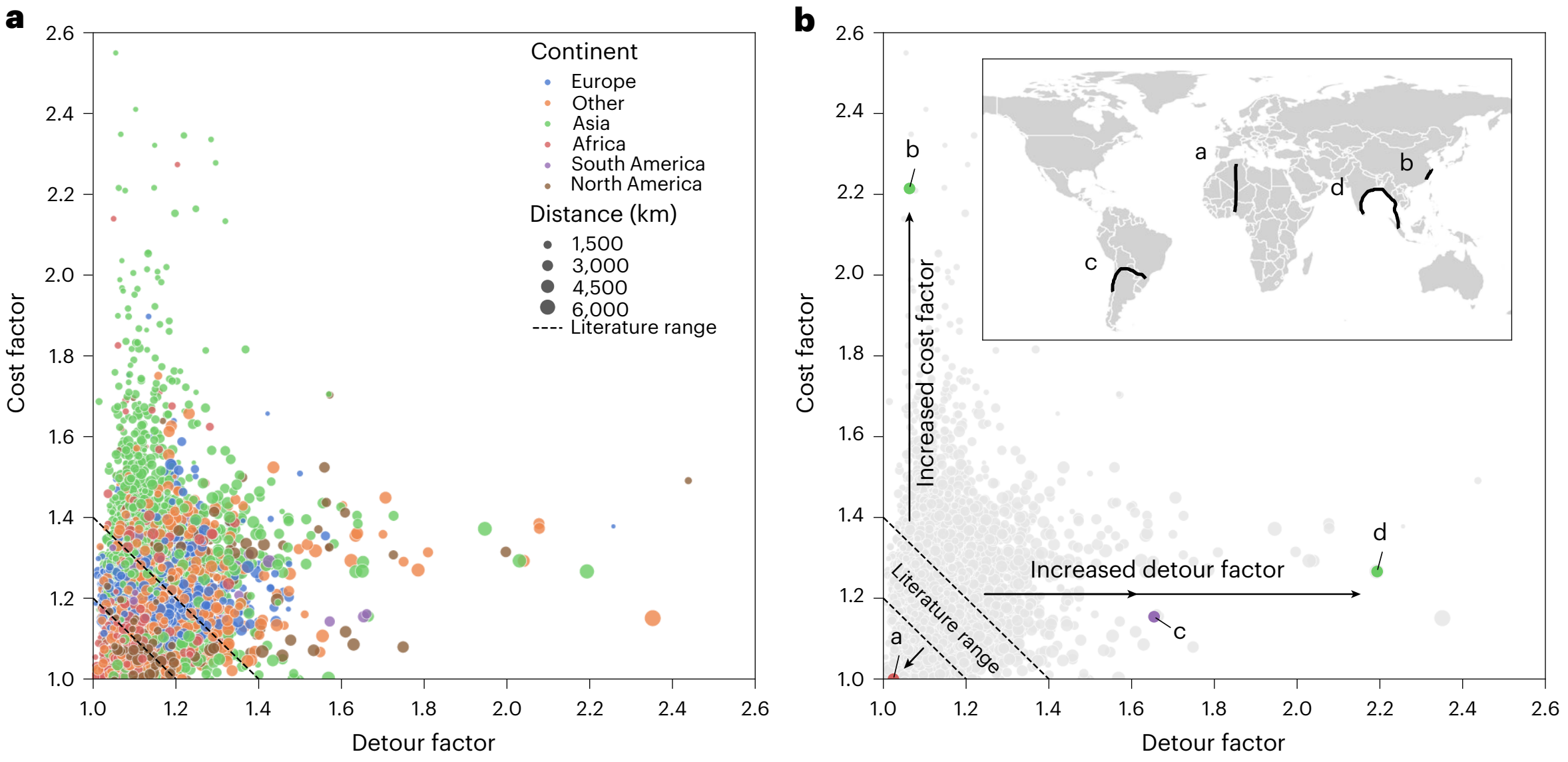

**Fig. 4 | Correlation between the detour and cost factors. a**, This figure presents a detailed analysis of the detour and cost factors, the two components of the total factor, for each global hydrogen pipeline route. Each route is represented as a bubble, with the size of the bubble indicating the distance of the route. The *x* axis shows the detour factor, calculated as the ratio of the actual pipeline distance to the straight-line distance, while the *y* axis shows the cost factor, reflecting increased costs due to geographical conditions. The dashed lines mark the standard literature range of 1.20–1.40, referring to the described standard method estimating pipeline costs using detour factors. The scatter plot forms an 'L' shape, illustrating that high total factors are primarily driven by either a high cost factor or a high detour factor but rarely both. Routes with higher detour factors tend to have greater straight-line distances compared with those with elevated cost factors. Notably, while many routes exceed literature values, a substantial number have total factors below these values, underlining the variability and uniqueness of each route. **b**, Highlighted examples 'a' to 'd' represent specific cases: 'a' shows a route with a very low total factor, 'b' shows a route with a high cost factor, and 'c' and 'd' show routes with high detour factors. These examples are further explored in Fig. 5. Basemap from Natural Earth (https://www.naturalearthdata.com).

Route b, connecting Fujian with Shanghai in China, also has a very small detour factor (1.06) from the straight line. However, in this case, the cost increase due to geography is high (cost factor of 2.21) caused by uneven terrain, mixed land use (cropland, forests and grasslands), offshore segments and occasionally densely populated areas along the route. This results in a total factor of 2.35.

By contrast, for routes c and d, the high total factors are driven by the detour and not the cost factor. Route c connects South Brazil and Chile, running through barren areas, the Andes at the Eastern Cordillera in Bolivia and through the Atacama Desert, a region with good hydrogen supply potentials. The route avoids more costly areas in Argentina and has a cost factor of only 1.15. However, this leads to a longer detour of 1.65, culminating in a total factor of 1.91.

For route d, connecting Sumatra, Indonesia, to Sri Lanka, the detour is considerably longer. The deep waters of the Bay of Bengal, reaching depths of more than 3,000 m, prevent the construction of an offshore pipeline along the straight-line route. Instead, a detour over the mainland to the north is necessary. The route traverses the ocean at shallow depths near Myanmar and Bangladesh, before reaching the Indian mainland. The protected areas in the north of Sri Lanka prevent a more direct route along the Indian shore. These obstacles result in a very high detour factor of 2.19, more than doubling the straight-line distance, which yields a total factor of 2.78 when combined with the cost factor of 1.27 including the offshore share.

## Country-specific WACCs further increase the range of costs

Figure 6 shows a stepwise approach to disentangle the effects of considering geographical and political factors compared with the conventional total factor approach. Step 0 (Fig. 6a) represents the current standard approach with a uniform detour factor of 1.30 and a uniform WACC of 6%, resulting in a single line, and no distinction made between different pipeline connections.

The first step (b) incorporates geographical characteristics, but the WACC remains at 6%. This results in a broader distribution of LCOT. This distribution as a whole can shift along the *y* axis depending on the WACC. The resulting routes are shown in Supplementary Fig. 1.

The second step (c) then includes country-specific WACCs, representing political factors, resulting in an even broader spread of LCOT. A box plot for all pipeline WACCs is shown in the top-right corner and more details on the country-specific WACCs are presented in Supplementary Table 1. Compared with the routes with a standard WACC, the range broadens at both ends, with the lowest LCOT slightly lower and the highest LCOT higher. In all three steps, the 'peak' LCOT stays the same at around 15 €/$MWh_{H2,LHV}$/$1000km_{straight\text{-}line}$, but the distribution spreads more with each step. The same LCOT of the histograms' peak can be explained by the fact that the lowest costs are based on a combination of favourable factors such as advantageous geography, existing infrastructure, low population density and a low WACC. If a constant WACC of 6% is applied to the same pipeline route, this leads to increased costs for this lowest-cost pipeline owing to the higher WACC compared with a lower country-specific rate.

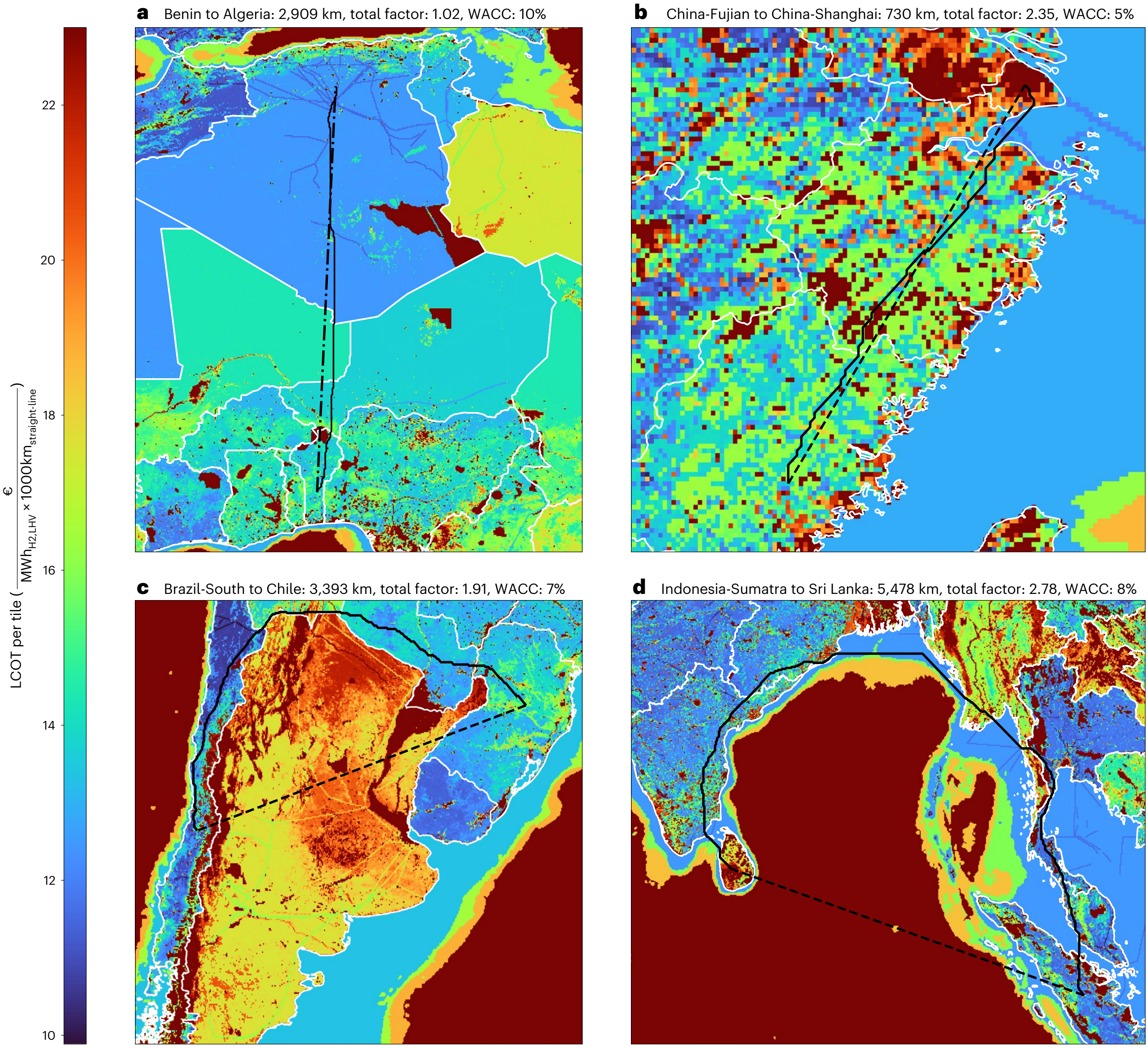

**Fig. 5 | Geographic overview of example routes. a–d**, This figure shows a close-up of the example routes from Fig. 4b, highlighting their geographical context with the LCOT of each tile shown in colour. Each map shows the calculated route (solid line) and the straight-line connection (dashed line) for comparison: the route from Benin to Algeria closely follows the straight-line owing to favourable geographical conditions leading to a low total factor of 1.02 (**a**). The route in China from Fujian to Shanghai also runs close to the straight-line but incurs a high total factor of 2.35 owing to more challenging geography (**b**). The route from South Brazil to Chile avoids costly areas in Argentina and East Chile, resulting in a higher detour factor, far from the straight-line, and a total factor of 1.91 (**c**). The route from Sumatra, Indonesia to Sri Lanka is forced to take a considerable detour, because the ocean is too deep to cross directly along the straight-line. This detour results in a total factor of 2.78 (**d**). Basemap from Natural Earth (https://www.naturalearthdata.com).

## Discussion

Our findings demonstrate that incorporating regional geographical and political characteristics substantially affects the estimation of hydrogen pipeline transport costs, with variations of up to a factor of three. This underscores the limitations of applying a universal factor for cost estimation, as it may lead to inaccuracies. Given that existing work on hydrogen imports to regions that will probably need them—such as Germany, the Netherlands, Japan and South Korea[41,42]—indicates that transport costs comprise only a small share of total hydrogen supply costs[11,18,40], the question arises whether the proposed detailed analysis is necessary on an international level.

To demonstrate the relevance of our findings, we analyse the effect of our calculations on the European energy system under five different scenarios, making use of the two well-established energy system models Enertile[8] and PyPSA-Eur[43] as well as existing scenario frameworks[43,44]. More specifically, we compare a base case using a detour factor of 1.2 with one using our route-specific cost calculations for each of the scenarios. The scenarios differ in hydrogen demand—ranging from a low demand of 540 $TWh_{H2,LHV}$, outsourcing much of Europe's energy intensive industry, to a high demand of 2,360 $TWh_{H2,LHV}$ assuming hydrogen demand from industry and long-distance heavy-duty freight transport. We adopt the general modelling assumptions and the input parameters of the energy system models from the existing scenario frameworks. With this energy system analysis, we aim to illustrate the practical implications of our results. We compare using route-specific pipeline cost with the traditional approach and do not perform a systematic analysis of other model parameters, and we do not validate these existing works (see Supplementary Note 1 for a detailed discussion of this aspect and its limitations).

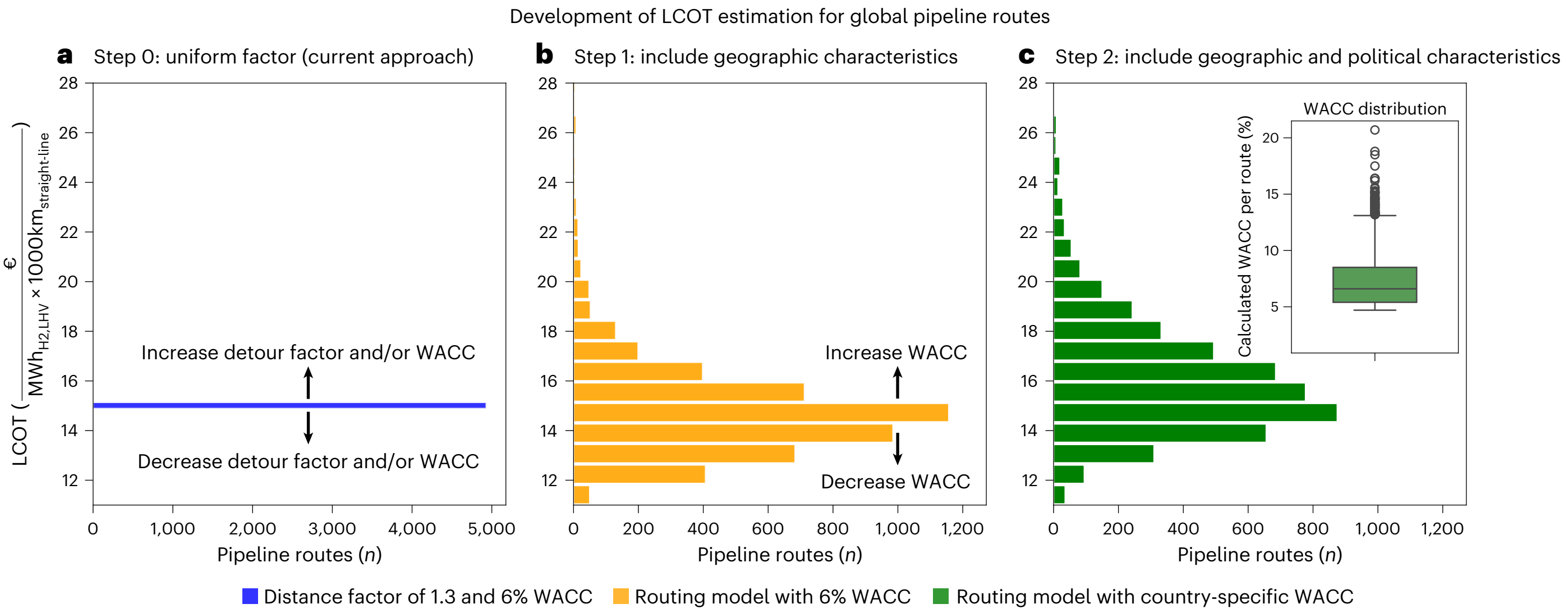

**Fig. 6 | Comparison of LCOT distribution between current conventional approach and adding geographical and political characteristics. a–c**, The figure shows how the LCOT of hydrogen develops by including increasingly detailed geographic and political factors in the modelling process. The current simplified approach is shown in step 0 (**a**), where all pipeline routes have one distance-specific cost, which can increase or decrease depending on the chosen detour factor or WACC. In step 1 (**b**), geographic factors such as land use, topography and existing pipeline infrastructure are integrated, while the WACC, representing political factors, remains constant. Including geographic factors broadens the distribution of LCOT, reflecting increased variability due to geographic influences. Step 2 (**c**) additionally incorporates country-specific political factors by varying the WACC, which further increases the distribution of LCOT. This step highlights the critical role of both geographical and political conditions in influencing pipeline costs. The top-right corner of **c** presents a box plot of the range of country-specific WACCs used. Inset: $n = 4{,}934$; minimum–maximum, 4.7–20.7%; median, 6.6%; Q1–Q3, 5.4–8.5%; whiskers, 4.7–13.1%; outliers, 71.

Figure 7 shows the relative changes in hydrogen flow for the used pipeline routes comparing our approach using region-specific cost factors with the traditional approach for each scenario. The absolute number of used pipeline routes increases from 12 in the E-540 scenario to 30 in the P-1800 scenario. In all scenarios, at least 83% (E-540) of routes were impacted (change of more than ±1%) by changing the cost assumptions with a maximum of 93% of routes impacted in the P-1800 scenario. In four of the five scenarios, fewer than 9% of pipeline routes show 'no change'. In the E-1600 scenario, half of all flows change by more than ±50%, and the flow direction reversed in one route. In this case, Austria switches from importing from the Czech Republic to exporting to it. In three out of the five scenarios, at least one route is completely discarded (−100% of hydrogen flow), and all except the scenario with the smallest hydrogen demand (E-540) have routes where the hydrogen flow increases by more than 100%.

These results underscore that, across different scenarios with different hydrogen demand levels, considering region-specific pipeline transport cost assumptions can substantially influence the amount and direction of hydrogen flows in an optimal energy system, even though transport costs only represent a small fraction of total hydrogen supply costs. The consistency of our general findings across all five scenarios suggests that the observed effects reflect structural responses to the use of route-specific pipeline cost inputs. Hence, such differentiations provide valuable insights into developing cost-efficient pathways towards a carbon-neutral energy system.

The findings of this study underscore the bandwidth of hydrogen pipeline transportation costs across different regions, which is much broader than typically reflected in current energy system analyses. Integrating route-specific geographical and political factors into estimating hydrogen transport costs reveals that the LCOT can vary substantially—by up to a factor of three—depending on the regional characteristics considered. Our analysis has four implications for energy system modelling and strategic decision-making for policymakers and industry stakeholders.

First, integrating region-specific pipeline costs into energy system models results in a more accurate depiction of the energy system of the future. An accurate analysis of the energy system is important for decision-making in industry and policy, for example, regarding infrastructure planning and investments, energy trade strategies, as well as respective policies, and should, hence, be considered by energy system modellers. Conventional models applying a single detour factor or cost estimate fail to capture the specifics of individual pipeline routes and may over- or underestimate the costs and trade flows between nations. This can result in misinformed decisions regarding infrastructure investments and energy trade strategies. Our datasets with the calculated global pipeline routes provide stakeholders with the data needed to perform more detailed and context-specific analyses, ultimately leading to more informed decisions in industry and politics.

Second and more specifically, our findings highlight the importance of considering both geographic and political factors when considering region-specific pipeline cost. Our analysis shows the substantial effect of both factors. We think that its importance increases as the global demand for hydrogen continues to grow and recognizing the complexities and disparities in pipeline transportation will be increasingly critical for assessments of the cost-effectiveness and efficiency of hydrogen solutions. Hence, and in addition to the effects mentioned above, these assessments are also important for supporting global decarbonization efforts and, therefore, for developing cost-efficient pathways towards a sustainable, net-zero energy system.

Third, a better understanding of the regional dynamics of hydrogen transportation costs can influence strategic decisions in both industry and policy related to the preferred way to transport hydrogen. In regions with lower-than-average pipeline costs, pipelines may have a clear advantage over alternative methods, such as shipping.

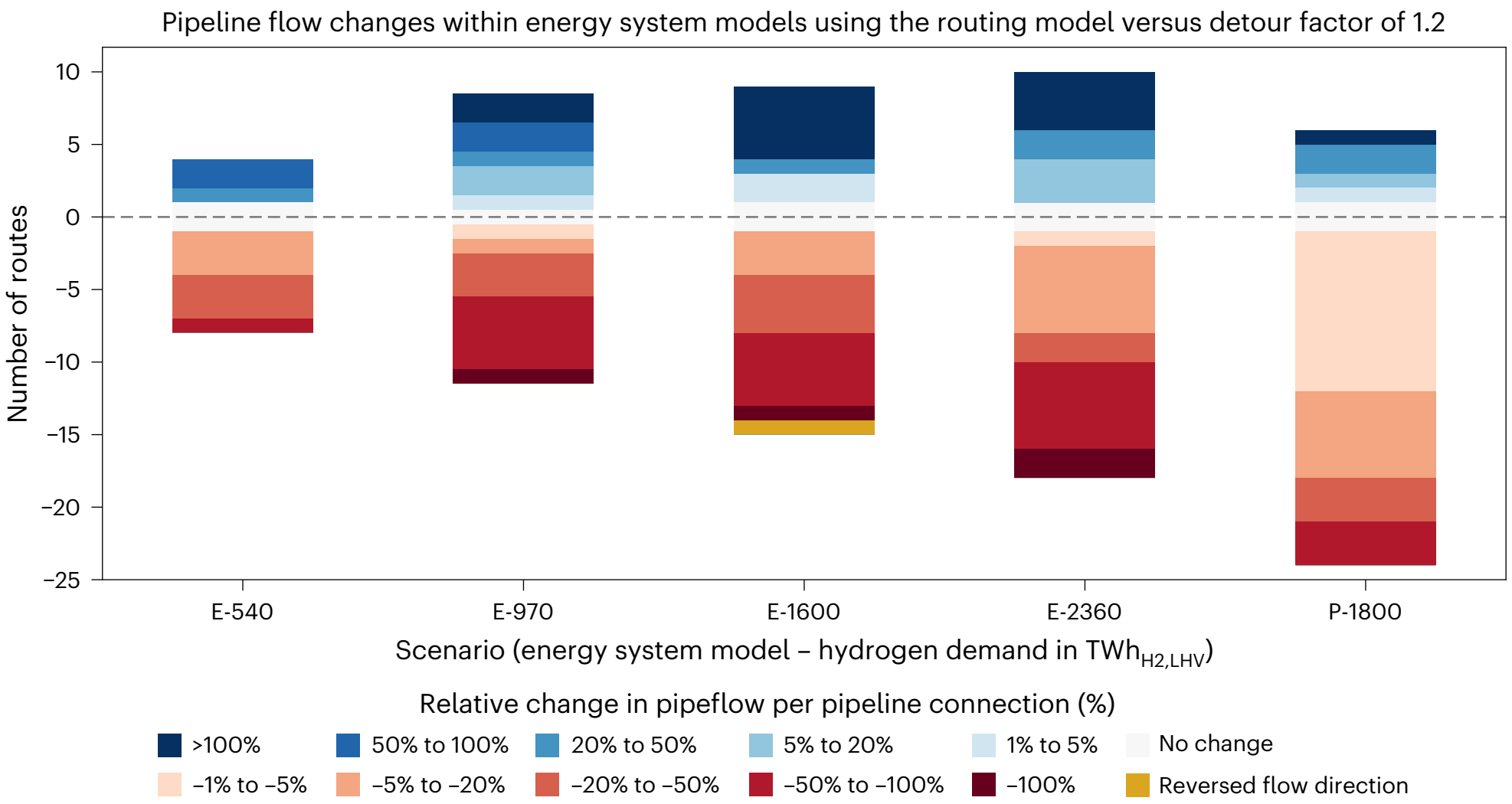

**Fig. 7 | Changes in amount and direction of hydrogen transported via each route comparing using region-specific characteristics with using a detour factor.** The figure shows the relative change in hydrogen flow per route for each of the considered scenarios (see Supplementary Table 7 for the corresponding values). The scenario names refer to the used energy system model Enertile (E) or PyPSA (P) and the scenarios hydrogen demand for Europe in 2050, rounded to 10 $TWh_{H2,LHV}$. Each bar shows the relative change in hydrogen flows for each route comparing the use of the route-specific costs applying the proposed approach with considering region-specific characteristics with applying a detour factor of 1.2. 'Reversed flow direction' means that the route's flow direction was reversed (for example, changing the flow from A to B in the base case to a flow from B to A) whereas 'no change' means the relative difference in the quantity of hydrogen transported along the route between the calculation using region-specific characteristics and the one using a standardized detour factor is within a range of ±1%.

Conversely, in regions with higher pipeline costs, there may be a shift towards shipping or alternative transport solutions. Industry players could plan their strategies accordingly and policymakers that want to support the development of the hydrogen economy can focus on financially attractive cases and potentially save public cost. Moreover, this understanding can be pivotal when establishing international trade relationships, as stakeholders may prefer regions where pipeline transport is financially more attractive.

Finally, we complement the work of this study by providing an online tool for calculating custom routes going beyond the analysis presented in this study to allow stakeholders to perform tailored analyses that can reflect their specific requirements and contexts. The practical application of the developed model can help decision-making in the context of the construction of hydrogen pipelines and support companies and government authorities in the planning of cost-effective hydrogen transport routes globally. In doing so, the model helps to identify potential bottlenecks or conflict points, such as densely populated areas or ecologically sensitive zones, at an early stage. The work therefore not only contributes to the scientific debate but also has the potential to have a direct and practical impact on the energy transition and the associated infrastructure design.

## Conclusion

This study provides insights into the regional variability of hydrogen pipeline transportation costs, highlighting the necessity of incorporating geographical and political factors into cost calculations and energy system modelling. Our work demonstrates that conventional approaches, which often rely on uniform detour factors and uniform distance-specific costs, fail to capture the complexities inherent in international hydrogen transport. By employing a GIS-based model that accounts for region-specific characteristics, we reveal that the LCOT can vary substantially—by up to a factor of three—depending on the unique conditions of each route.

By applying our results to five scenarios and using two different well-established European energy system models, we show that the regional variations in LCOT substantially influence hydrogen trade flows in the optimal energy system, even if pipeline transportation costs only account for a small share of total hydrogen supply costs.

Our findings indicate that a more nuanced understanding of hydrogen infrastructure is essential for effective energy system modelling and investment decisions. A higher level of granularity not only increases the accuracy of energy system models but also provides valuable insights for policymakers and stakeholders involved in the planning of and investments in hydrogen infrastructure. Our findings also suggest that strategic decisions regarding hydrogen transport should consider these regional disparities to optimize the economic viability and efficiency of future energy systems.

Future research should address the limitations identified in this study, particularly regarding the potential for considering an interplay of new and repurposed existing pipelines and the influence of political relationships on international trade dynamics. By expanding the dataset and approach, possibly integrating it into a network topology optimization[45,46] that considers shared pipeline use among multiple hubs, future studies can continue to hone our understanding of a potential future hydrogen system and contribute to the development of a robust, decarbonized energy system. Ultimately, this research lays the groundwork for more informed decision-making in the rapidly evolving landscape of hydrogen as a key energy carrier in the transition to a sustainable future.

## Methods

This Article introduces an approach to determining the routes and costs of hydrogen pipelines on a global scale by considering regional characteristics (see Supplementary Fig. 2 for an overview). More specifically, we consider region-specific (1) geographical and (2) political factors on a global scale. This section provides an overview of the method. The geographical factors distinguish 11 types of land use, protected areas, topography, bathymetry, population density and existing pipeline corridors, which consider different levels of existing and planned pipeline infrastructure for oil and gas. The political factors consider economic, financial, social and political stability, as well as legal (un)certainty, and are represented by a country-specific WACC, incorporating existing country-specific equity risk premiums[47]. Our analysis considers both onshore and offshore pipelines, ensuring a holistic evaluation of potential routes. We obtain the hydrogen pipeline transport cost in two steps: first, we calculate the route and then the LCOT of hydrogen. The cost factor, which represents the individual increase or decrease of cost for constructing a hydrogen pipeline, is aggregated from each considered geographical factor for each tile on the global grid. The weight of each edge connecting the tiles is calculated, creating a global network graph (Supplementary Fig. 3). We use the Dijkstra algorithm to calculate the cost-optimal pathway through this network graph on the basis of a given start and end location. The first step results in the route, the route's average cost factor and WACC, the latter two weighted by the individual edge distance of each step. Second, we calculate the LCOT in €/MWh$_{H2,LHV}$ for each route supplementing the results of the route calculation (step 1) with techno-economic assumptions on the size and operation of the pipeline. We assume a uniform diameter for all pipelines to ensure comparability in our analysis. For real projects with a specific required throughput, the diameter and therefore also the LCOT can differ. While the cost factors directly affect the pipeline construction cost, we base our analysis on LCOT because they show the effect on the cost of the transported hydrogen. We further normalize the LCOT to 1,000 km of straight-line distance, that is, €/MWh$_{H2,LHV}$/1000km$_{straight\text{-}line}$, to ensure comparability to the current standard approach in literature. All cost assumptions are based on the reference year 2020, and all energy quantities refer to hydrogen's LHV. Each route's detour factor is calculated by dividing the route distance by the straight-line distance. In addition, we calculate the total factor, that is, by multiplying the detour and the cost factor, to represent the total increase in the standard distance-specific investment of a pipeline between a start and end point.

## Considered geographic and political characteristics

This section outlines the considered geographical and political characteristics impacting pipeline construction costs. These characteristics include land use, excluded areas, topography, bathymetry, population density, existing pipeline infrastructure and WACC. A cost factor is assigned to each characteristic, representing its influence on construction costs. The WACC is an exception, as it reflects financing costs rather than direct construction costs and is incorporated via the annuity factor (equation (1)).

## Geographic characteristics

As shown in Supplementary Fig. 3a, the spatial datasets for each characteristic are aggregated onto a global grid with approx. 12.7 million tiles, initially developed as part of the 'Enertile'[8,48] energy system model. The grid structure with tiles of a size of 42.5 km$^2$ serves as the foundation for projecting the necessary data. Supplementary Fig. 3b shows the final cost-factor aggregation on the global grid. In brief, land use is distinguished into 11 categories, with cost factors ranging from 1.0 to 5.0, and tiles containing more than 50% of protected land are assigned a cost factor of 99 to prevent crossing. Topography cost factors range from 1.0 for flat terrain to 4.0 for steep alpine slopes. Offshore segments are classified into five ocean depth categories, with depths exceeding 3,000 m being excluded. Areas with existing pipeline infrastructure receive a cost discount. Population density increases the cost factor above 250 inhabitants per km$^2$, reaching exclusion of the area (cost factor of 99) above 4,000 inhabitants per km$^2$. The cost factors for each category are presented in Supplementary Table 2, with the underlying data sources presented in Supplementary Table 3. A description of each characteristic is provided in Supplementary Note 2. The land usage, including restricted areas, and the topography were already preprocessed and rasterized during the development of Enertile. For more details on this, refer to refs. 49,50.

## WACC

The WACC is often used to estimate financing costs in (trans)national hydrogen transport, with many studies assuming a constant value of 4–6%[51]. This assumption implies a uniform investor risk across national borders, which in some cases can lead to an underestimation of the actual risks[31,51].

To include the political, economic, financial and institutional risks of the countries traversed in the routing, we use country-specific WACC values from Damodaran[47]. These are composed of a country-independent rate of 4.72%, reflecting the equity risk of a mature market, and a country-specific risk premium that accounts for macro risk factors such as economic, financial, social and political risks as well as the legal certainty of a country. Factors at the meso level, such as market and technology maturity or the quality and availability of resources, as well as project-specific aspects are not taken into account. Although this approach enables a more realistic representation of country-specific capital costs, it neglects advantages such as lower labour costs or simplified planning procedures, which are prevalent in some emerging and developing countries[52]. The WACC is assigned on the basis of administrative country borders, with a 30% area threshold for tile allocation. This method enables assigning a WACC even in complex border regions, such as three-country corners. If one tile lies within multiple countries, the country with the highest share is allocated. Offshore tiles that do not fall under any national jurisdiction are not assigned a specific WACC. They adopt an average WACC from the pipeline's start and end countries. Supplementary Table 1 presents the country-specific WACC on a global scale.

## Scenario setup parameters

To evaluate regional differences influenced by the introduced geographic and political characteristics, we calculate synthetic hydrogen pipeline routes. This requires techno-economic base parameters on the construction and operation of hydrogen pipelines, along with the designation of regional hubs, serving as connection points for each region.

## Distance-specific pipeline cost and base cost

To estimate the total investment required for a pipeline project, we estimate the base cost per kilometre of pipeline, which serves as the foundation for additional cost adjustments based on various characteristics. This estimation is derived from a comprehensive literature review that includes various cost functions, which are presented in Supplementary Fig. 4 and Supplementary Table 4. For consistency, all functions have been standardized to euros per kilometre, with values originally in other currencies converted and adjusted to the base year of 2020. An average cost function was derived to mitigate variability. This average function serves as the basis for subsequent cost calculations and provides a balanced representation of the compiled data.

Additional to the base cost function, all relevant operating and economic assumptions are summarized in Supplementary Table 5. The fixed annual maintenance and operating costs amount to 5% of the total investment. In addition, the variable operating costs for the compressors are derived on the basis of the amount of energy required for recompression from 30 to 100 bar, a general electricity price and a recompression interval of 250 km. With regard to the operating parameters of the pipeline, the detailed pressure loss and operating analysis by Krieg[16] is used, according to which substantial pressure losses in the pipeline can be ruled out and regular operation can be guaranteed.

## Country hubs

For global pipeline cost comparison, we calculate routes connecting 264 regions, assigning a regional hub to each region, serving as connection points (Supplementary Fig. 5). This work does not pre-assess locations for hydrogen supply or demand suitability, avoiding bias in hub location selection. Instead, a heuristic is applied universally without prior assumptions about a country's potential role as a hydrogen exporter or importer. Each hub's location is determined by weighing the region's area centroid at 20% and its population density centroid at 80%. This ensures that hubs lie in some proximity to civilization, while avoiding locations at the extreme edges, which is crucial for geographically complex and large countries such as Canada, Norway and Russia, where infrastructure is sparse and the area centroid may be far from developed and populated areas.

To account for their size and complexity, the countries USA, Russia, India, Brazil, Australia, Canada, Indonesia, Japan, Malaysia, Philippines, Vietnam and China are divided at a subcountry level, as also done in Brinkerink et al.[39], one of the few existing studies that does energy system modelling on a global scale and has investigated the effects of spatial resolution when doing so.

## Model formulation

The model is structured into two sequential calculations. First, the route optimization, where the least-cost route between two region hubs and its properties are determined using the global grid. Subsequently, in the cost calculation, the operating parameters are included to calculate the levelized cost of transporting hydrogen.

**Route optimization.** The geographic problem is mathematically modelled by representing each grid cell as a unique node *N* within a graph *G*, identified by an Area-ID and linked to specific cost factors. Nodes are connected by edges *E*, with the node's position determined by the centre of each cell. The weight of an edge $w(E)$ considers several factors: distance ($d$), topography cost factor ($f_t$), land use cost factor ($f_l$), population density cost factor ($f_p$), sea depth (bathymetry) cost factor ($f_b$), pipeline corridor cost factor ($f_c$) and the annuity factor (AF), that comprises the WACC and the lifetime ($n$) serving as the depreciation period. In cases where a tile cannot be assigned to a specific country, such as in international waters, the individual route-specific average values of the start and end point are used to calculate a WACC. The edge weight is calculated as follows:

$$w(E) = d(E) \times \big(1 + (f_t(E) - 1) + (f_l(E) - 1) + (f_p(E) - 1) + (f_b(E) - 1) + (f_c(E) - 1)\big) \times \mathrm{AF}(E), \tag{1}$$

with annuity factor AF,

$$\mathrm{AF}(E) = \frac{(1 + \mathrm{WACC}(E))^n \times \mathrm{WACC}(E)}{(1 + \mathrm{WACC}(E))^n - 1}, \tag{2}$$

where $d$ is the edge distance, $n$ is the number of years, WACC is the WACC, $f_t$ is the topography (slope) cost factor, $f_l$ is the land-use cost factor, $f_p$ is the population-density cost factor, $f_b$ is the bathymetry cost factor and $f_c$ is the existing pipeline corridor cost factor.

Given starting point a and end point b, the Dijkstra algorithm is used to calculate the least-cost pathway $P(\mathrm{a,b})$ within the network graph *G* by minimizing the sum of the weights of the required edges.

$$\min_{P(\mathrm{a,b}) \in G} \sum_{E \in P(\mathrm{a,b})} w(E). \tag{3}$$

The results include the path of the specific route, the total factor representing the distance-weighted average of all cost factors, the detour factor indicating deviations from the direct path and the route-specific WACC.

## Cost calculation

The LCOT of hydrogen is determined by combining capital expenditures (CAPEX), fixed operation and maintenance costs ($\mathrm{OPEX_{fix}}$) and variable operation cost ($\mathrm{OPEX_{var}}$). The CAPEX incorporates the calculated route-specific parameters total factor, annuity factor and route distance along with the base investment of the pipeline that depends on its diameter. The $\mathrm{OPEX_{fix}}$ is set as 5% of CAPEX and the $\mathrm{OPEX_{var}}$ includes the electricity costs for the compressor stations. Instead of using the LCOT over the individual route distance ($\mathrm{LCOT_{route\ distance}}$), we normalize the LCOT by 1,000 km of straight-line distance ($\mathrm{LCOT_{straight\text{-}line}}$), allowing comparison of both among different routes and with standard literature estimates for hydrogen pipeline costs, which also use straight-line distances. In the main text, this $\mathrm{LCOT_{straight\text{-}line}}$ is referred to simply as LCOT for conciseness. A consistent medium pipeline diameter of 0.9 m is used across all pipeline connections to ensure comparability. Hence, for real projects with a specific required throughput, the diameter and therefore also the LCOT can differ.

$$\mathrm{LCOT_{route\ distance}} = \frac{\mathrm{CAPEX} + \mathrm{OPEX_{fix}} + \mathrm{OPEX_{var}}}{\dot{Q}_{\mathrm{pipe}}(D_{\mathrm{pipe}}) \times \mathrm{flh}}, \tag{4}$$

with

$$\mathrm{CAPEX} = C_{\mathrm{base\ investment}} \times f_{\mathrm{total}} \times d_{\mathrm{route}} \times \mathrm{AF}, \tag{5}$$

$$\mathrm{OPEX_{fix}} = \mathrm{CAPEX} \times C_{\mathrm{O\&M,rel}}, \tag{6}$$

$$\mathrm{OPEX_{var}} = \frac{d_{\mathrm{route}}}{d_{\mathrm{recomp}}} \times p_{\mathrm{el}} \times E_{\mathrm{el,comp}} \times \dot{Q}_{\mathrm{pipe}}. \tag{7}$$

For comparability of results, the LCOT of each route is normalized by straight-line distance,

$$\mathrm{LCOT} = \mathrm{LCOT_{straight\text{-}line}} = \frac{\mathrm{LCOT_{route\ distance}}}{d_{\mathrm{straight\text{-}line}}} \times 1{,}000, \tag{8}$$

where $D_{\mathrm{pipe}}$ is the pipeline diameter, $\dot{Q}_{\mathrm{pipe}}$ is the pipeline capacity, flh is the full load hours, $C_{\mathrm{base\ investment}}$ is the base investment for the pipeline (Supplementary Fig. 4), $f_{\mathrm{total}}$ is the total cost factor, which is calculated from the individual factors in the route optimization, $C_{\mathrm{O\&M,rel}}$ is the operation and maintenance cost relative to CAPEX, $E_{\mathrm{el,comp}}$ is the electricity consumption for recompression, $d_{\mathrm{route}}$ is the route distance, $d_{\mathrm{straight\text{-}line}}$ is the route's straight line distance connecting the start and end point, $d_{\mathrm{recomp}}$ is the recompression interval distance and $p_{\mathrm{el}}$ is the electricity price. Supplementary Fig. 6 shows an aggregated LCOT calculated individually for each tile in the global raster.

## Sensitivity analyses

To assess the robustness of the LCOT distribution across uncertainty of input parameters, we conduct sensitivity analyses on both the political and geographic model inputs. We uniformly increase the WACC by +1%, +2% and +3% across all countries to represent a plausible range of project-specific risk premiums for hydrogen pipelines, which we did not consider in the base case. For the geographic cost factors, we separately increase the cost factor groups land use, slope, population density and sea depth by 20% for all values above 1 (cost factors that represent a cost increase relative to the base cost). For each sensitivity, we recalculate all routes globally and compare the resulting LCOT distribution against the base case showing our main results (Supplementary Note 3, Supplementary Fig. 7 and Supplementary Table 6).

## Assumptions for energy system modelling

To demonstrate the relevance of using regional pipeline assumptions in an energy system model, we analyse the effect of replacing commonly used detour factors by using our routing-model results as input to European energy system models. We use existing scenario frameworks for the two well-established energy system models Enertile[8] and PyPSA-Eur[43] and consider five scenarios for 2050. In doing so, we test the impact of incorporating our calculated region-specific hydrogen pipeline routes over a range of very different hydrogen demand scenarios, as well as using different energy system models. More specifically, we compare the resulting hydrogen pipeline flows of using our region-specific routes, with those obtained using the standard detour factor approach for each of the energy system model scenarios.

The scenario names consist of the model's name and the hydrogen demand in $TWh_{H2,LHV}$. The four Enertile scenarios (E-540 to E-2360) are based on Fleiter et al.[44], and the PyPSA-Eur scenario (P-1800) is based on Neumann et al.[43]. Further scenario descriptions are provided in Supplementary Note 4.

For each of the scenarios, we run two energy system model runs where in one, the hydrogen pipeline transport cost was calculated using a detour factor of 1.2 and in the other one, the pipeline transport cost was calculated using the results discussed in this work. We harmonize the spatial resolution of the two different models to national level and apply our calculated route-specific hydrogen pipeline detour and cost factors. The results are shown in Fig. 7.

The models are well established, and the input parameters have not been changed from the original sources, which is why we consider further validation of the models and their method as beyond the scope of this work. Nevertheless, the original study by Neumann et al.[43] conducted extensive sensitivity analyses (see Supplementary Note 1 for details).

## Data availability

All resulting pipeline data including costs and geographic routing information can be accessed via Zenodo at https://doi.org/10.5281/zenodo.15308870 (ref. 53). Source data are provided with this paper.

## Code availability

The developed model HyTROM (Hydrogen Transport Route Optimization Model) can be accessed via a graphical user interface implemented using streamlit at https://websites.fraunhofer.de/hytrom/. The user interface enables the calculation of any potential pipeline route, including those analysed in the paper, and even enables users to compute new custom routes anywhere in the world using either our techno-economic assumptions, other assumptions from literature or their own.

## Acknowledgements

We thank K. Franke, J. Fragoso and B. Lux for the valuable discussions and for providing the detailed datasets from their publications[49,50], which were crucial for the success of this work. During the preparation of this work, the authors used DeepL to improve the clarity of the English writing as well as ChatGPT 5 and Claude Opus 4.6 to enhance the readability and language of the paper. After using this tool/service, the authors reviewed and edited the content as needed and take full responsibility for the content of the published article.

## Author contributions

B.W. suggested the initial research question, which A.S. and R.M. extended. L.K. and B.W. collected the data and developed the model. B.W. performed the analyses and created the Figs. 1–7. All authors interpreted the results. B.W. and A.S. conceptualized the study, and B.W. wrote the study with contributions from L.K. and A.S. All authors reviewed and edited the study.

## Funding

B.W. received funding for this research work within the projects HYPAT (grant no. BMBF FKZ 03SF0620A) and HySecunda (grant no. BMBF FKZ 03SF0734), both funded by the German Federal Ministry of Education and Research (BMBF), and through the Fraunhofer Cluster of Excellence 'Integrated Energy Systems' (CINES).Open access funding provided by Fraunhofer-Gesellschaft zur Förderung der angewandten Forschung e.V.

## Competing interests

The authors declare no competing interests.

## Additional information

**Supplementary information** The online version contains supplementary material available at https://doi.org/10.1038/s41560-026-02141-3.

**Correspondence and requests for materials** should be addressed to Bastian Weißenburger.

**Peer review information** *Nature Energy* thanks Tubagus Aryandi Gunawan, Michael Moritz and Lucas Sens for their contribution to the peer review of this work. Peer reviewer reports are available.

**Reprints and permissions information** is available at www.nature.com/reprints.

**Publisher's note** Springer Nature remains neutral with regard to jurisdictional claims in published maps and institutional affiliations.